\documentstyle[preprint,aps] {revtex}
\begin{document}
\draft
\title{On the Shape of the First Collapsed Objects}

\author{S.F. Shandarin, A.L. Melott, K. McDavitt, J.L. Pauls, and J. Tinker}
\address{Department of Physics and Astronomy, University of Kansas,
Lawrence, Kansas 66045} \date{\today} \maketitle
\begin{abstract}

Since the early seventies, there was a conjecture that the first
collapse of a selfgravitating dust--like medium (appropriate
approximation for nonbaryonic dark matter) results in the formation of
a ``pancake" object, that is a thin surface. The conjecture has been
based on the Zel'dovich approximate solution of the nonlinear
gravitational instability of a generic smooth density
perturbation. Recent works cast doubt on the Zel'dovich conjecture,
suggesting that the first collapse might be point--like or
filament--like rather than pancake--like. Our $N$--body simulations
show first pancake collapse. We can reject with 97\% confidence the
Bayesian prior that the other kinds of collapse are more or equally
probable.

\end{abstract}
\vskip .25in
\newpage
\section{INTRODUCTION}

In the evolution of gravitational clustering in the expanding
universe, it has gradually been recognized that the first collapse is
usually anisotropic. This has important consequences on all scales,
from the formation of stars, galaxies, or superclusters. Although it
may be visible today only in superclusters, which are just collapsing
now, it has consequences for the formation of all objects by gravity,
assuming mathematically generic initial conditions without special
symmetry and neglecting pressure gradients.

A schematic history of the question follows: The known exact solution
obtained for the spherically symmetric, non--rotating, pressure--free
case, (e.g.  [1] ) predicts two types of collapse from rest. If the
initial density is monotonically decreasing then the first collapse is
point--like toward the origin. However, if the initial density is
non--monotonic then the first collapse is shell--like. Considering a
uniform, non--rotating, pressure-free spheroid Lin et. al. [2] found
that it collapses either toward a disk or a spindle depending on
whether it is oblate or prolate at the initial time.

Zel'dovich\cite{zel} proposed an approximation for a generic initial
perturbation which predicts that the first collapsed objects have a
pancake--like shape. Gurevich and Zybin [4] revisited the issue and
concluded that the nondissipative gravitational collapse of a generic
perturbation results in the formation of a stationary dynamical
structure with a point--like singularity at its core $\rho\propto
r^{-24/13}$.  Recently, Bertschinger\cite{bert} and Bertschinger and
Jain\cite{bert94} proposed a purely local gravitational instability
solution based on General Relativity which implied prolate collapse to
filaments comes first. This would have strong implications for star,
galaxy, and supercluster formation, if it is also true in weak
gravitational fields inside the horizon. Kofman and Pogosyan\cite{kof}
and Bertschinger and Hamilton\cite{bertham} showed that this solution
had neglected certain terms of the same order as others included in it
which may be justified in ultrarelativistic cases, but not in the
Newtonian limit. The difference between the collapse in a dust--like
matter in Newtonian and ultrarelativistic cases was stressed in
Zel'dovich and Novikov\cite{zn}; see also Matarrese, Pantano and
Saez\cite{mps}.  This provides a renewed justification for the
neo--Newtonian approach generally used for studying low--amplitude
cosmological perturbations inside the horizon.

However, as noted by Bertschinger and Hamilton\cite{bertham} it does
not resolve the question whether pancakes or filaments form
first. Although the Zel'dovich\cite{zel} approximation (ZA) predicts
pancakes, this approximation is not exact in three dimensions. It is
known, for example, that collapse in nonlinear gravitational
clustering simulations proceeds faster than ZA predicts. It is
therefore important to determine whether the quasi--two--dimensional
structures predicted by ZA really occur.

One should distinguish between two statements we might make: (1) The
first collapse is always pancake--like. (2) The first collapse is
usually pancake--like but could be filament--like in some special
cases. In this paper, we present evidence for the second (weaker)
statement based on numerical simulations.

The initial conditions we set up are of a generic type, which means
that a smooth small arbitrary perturbation does not change
qualitatively the type of initial condition in any sense.

\section{COMPUTATIONS}

We examined an ensemble of five $N$--body simulations on a $128^3$
Particle--Mesh gravitational clustering code with periodic boundary
conditions.  Initial conditions are impressed by the now--standard use
of a random number generator to create choices of phase and
Gaussian--distributed amplitude for various Fourier components of the
initial density fluctuations, and the motions in response to these. At
the very low amplitudes the ZA we used and Eulerian linear
perturbation theory of the growing mode are essentially
indistinguishable. We also stress that using shot noise with a dying
mode component [11] or a logarithmic distribution of modes [12] has
not made noticeable differences.  Further details on simulation
methods can be found in \cite{mel93}. The initial conditions
corresponding to the growing mode were constructed in four
realizations with initial fluctuations of wavenumber 1 through
$\sqrt{3}$ in units of the fundamental mode of the box. Thus, the
minimium wavelength present in initial conditions is 74 mesh units.
Although we wished to examine the first collapse in detail, a smaller
upper bound on wavenumber would cause alignment with coordinate
axes. An additional simulation with initial wavenumber range 1 through
3 (minimum wavelength 43 mesh units) was performed as a check (\#1 in
Table I). We found nothing special in this case.  All simulations were
started with $rms$ density fluctuation $\sigma \sim 0.03-0.04$ in
order to allow time (an expansion factor of $\sim 17$) for transients
to die out and full growing mode including nonlinear effects to
establish itself. Two simulations (\#2 and \#4), as a check, started
with half the initial amplitude and ran for twice the expansion
factor. We found no particular difference between runs with different
amplitudes.  At the initial stage the density perturbations look like
ordinary three--dimensional smooth Gaussian fields. The structures
shown below are resulted from nonlinear growth of the density
fluctuations due to gravitational instability.

All previous studies of the collapse of a smooth perturbation suggest
that the trajectories of the particles are smooth prior to the
shell--crossing (see e.g.  [1,9]). We have found no evidence against
this assumption. We also have found no evidence that the particles
might bunch up thickly without undergoing following
shell--crossing. Thus we follow Zel'dovich and define the first
collapsed objects as the regions where the first shell--crossing
occurs. Formally this definition does not assume that the particles
undergone shell--crossing form a gravitationally bound object, though
it is likely at the later stages.

We stopped the simulations after the {\it first} shell crossings. Our
timesteps are very strictly constrained so that the {\it fastest}
particle could travel 0.4 grid cell (out of 128) in a single
timestep. All particles were tagged which had local shell crossing (as
determined by whether the local volume element had gone negative).
Thousands of particles typically shell crossed for the first time in a
single step. We then examined the distribution of these particles. It
is worth stressing that the particles in question show the regions
between caustics and do not represent well the density distribution.
They were all highly anisotropic and resembled surfaces rather than
lines. One was ribbon--like but still essentially very flat. We will
illustrate this with multiple pictures from one simulation; other
simulations look similar.

Figure 1 shows three orientations of a typical surface (\#5 in Table I)
viewed along the three eigenvectors of the initial deformation
tensor toward the middle of the surface. Figures 1a and 1b suggest
finite thickness ($\sim 1$ to 6) but this is because the surface is
curved (bowl--like). In Figure 2 we show two cross sections of this
surface to indicate its thinness. Figure 1c shows the pancake region
face on. The reader should not be misled by the little rows of
particles which are the usual result of the standard ``quiet start"
with particles on a slightly deformed cubic lattice.

Figure 2a shows a cross section perpendicular to the $z$--axis and
Figure 2b shows a cross section perpendicular to the $y$--axis. Both
cross sections are very thin. They suggest that the real thickness of
the region ($\sim$ the distance between caustics) is of order one mesh
unit, while the diameters (size in $y$ and $z$ directions as seen in
Fig. 1a) are about 37 and 17 mesh units. From this we conclude that
the shape of the region is pancake--like with approximate ratios
1:17:37, rather than filament--like. We looked closely at many more
additional thin slices and concluded that the actual maximum thickness
was always $<0.8$, in agreement with the timestep constraint. The
other dimensions were much larger, as can be easily seen.

Catastrophe theory suggests that the diameters of a pancake grow as
$\sim(t-t_c)^{1/2}$ and its thickness (defined as the distance between
caustics) as $\sim (t-t_c)^{3/2}$, therefore the ratio of the
thickness to the diameter is proportional to $\sim(t-t_c)$ at small
$t-t_c$ (here $t_c$ is the time of the formation of the first
singularity)\cite{arn82}. Also the diameters are not equal to each
other in a generic case. In our simulation we plot Figures 1 and 2
after a small but finite time from the first local crossing (the first
``singularity" to the accuracy of the simulation) and therefore expect
small but finite thickness of the pancake.

In contrast to Figure 2, Figure 3 shows {\it all} particles in thin
slices orthogonal to the principal axes of the initial deformation
tensor at the largest eigenvalue. One can easily see the difference
between the density distributions (Fig. 3) and the shape of the
collapsed region (Fig. 2).  All the statements about the shapes of the
first collapsed regions derived from ZA refer to the shapes of
collapsed regions (Fig. 1 and 2) which may be similar but not the same
as the density distributions (Fig. 3) especially if the resolution is
not sufficiently good. We believe that it is worth keeping in mind
this difference while analyzing the results of $N$--body simulations.
\section{CONCLUSION}

Our simulations suggest that (to the limit of their accuracy) the
first stage of collapse of a generic gravitational system is usually
to a thin sheet as suggested by ZA.  (Obviously we cannot say anything
about the evolution of shapes between the last ``uncollapsed" and the
first ``collapsed" stages, but we stress that our timesteps are
shorter relative to the characteristic formation time of the stuctures
under consideration than any simulations to date.) This should be
taken into account in all gravitational instability theory from star
formation through large--scale clustering. Superclusters, now
experiencing their first collapse, should include sheetlike
structures. Filaments (another type of generic structure [14]) may be
easier to see due to their higher density constrast and possible gas
cooling effects (J.P. Ostriker, personal communication), but our
results indicate they will be second generation objects formed by
flows inside sheets.

In the presence of small--scale perturbations in the initial spectrum
(which is the most likely case in cosmology) these pancake--like
structures are not as smooth as the pancakes discussed in this
paper. As we mentioned before, there is a theoretical question
concerning the type of the first collapse in a dust--like medium. Our
results should not be interpreted as totally excluding the possibility
of the first collapse to filament--like structures. It is well known
from second order perturbation theory that the rate of collapse along
one principal axis depends on the rates of collapse along the other
principal axes. In principal, it may change the type of the collapse
in some cases. On the other hand, the general solution with the
maximal number (eight) of physically arbitrary functions of three
variables in a dust like medium suggests gravitational collapse is
pancake--like \cite{{lif},{gris}}.  In Katz {\em et al}\cite{kat} it
is stated that ``the first objects form in filaments from almost
two--dimensional collapses in agreement with the approximate analytic
theory of Bertschinger and Jain," which appears to contradict our
results.  We did not investigate all options for Gaussian initial
conditions.  Our initial conditions were particular random
realizations of Gaussian initial conditions, with formally $k^{-1}$
power spectrum of density fluctuations in the range of $k_f\leq k\leq
\sqrt{3}k_f$ (or $3k_f$ in one case).  But we would like to stress
that they were of a mathematically generic type.

We have presented in detail the results of the simulation of one
pancake.  However, we totally studied five realizations of the initial
conditions. All showed similar pancake--like structures (sometimes
elongated); see Table 1 where we list the thickness, width and length
for the first collasped objects in our first five realizations.  We
find neither a single filament--like collapse in our simulations,
(filaments would be expected on the basis of the hypothesis of
Bertschinger and collaborators[5,6]) nor point--like collapse [4].  If
we use a prior hypothesis that the ZA and HA descriptions of first
collapse are equally probable, we can reject this on the basis of our
experiments with 97\% confidence. Alternatively, we may assume
pancakes and other structures form with some probabilities and try to
estimate that probability. A sequence of five pancakes would be more
probable than the sum of all other sequences' probabilities if the a
priori probability of a pancake were 87\%.  We note that our objects
are all smaller (much smaller in thickness) than the minimum
wavelength in the initial perturbations, and thus represent the first
generation of collapsed objects. Objects formed on any scale in
hierarchical clustering $N$--body simulations, such as those of Katz
{\it et al.}\cite{kat}, are larger than the Nyquist wavelength of the
initial spectrum, and therefore a later generation and irrelevant to
the question studied here. However, we do comment that such
simulations might be expected to show one--dimensional collapse on the
objects where the things are just becoming mildly nonlinear.
Recently, observational evidence has appeared to suggest there are
sheetlike neutral hydrogen clouds at moderate redshift [19].
Quantitative evidence for pancake--like morphology for such objects
(as well as filaments existing at later stages of dynamical evolution)
has been found in hierarchical clustering simulations\cite{bab}.
However, this technique does not measure the distance between
caustics, discussed in this paper, and does not take into account the
thin bowl--like shape of the first pancakes.  The formation of the
filament--like structures as well as compact clumps of higher density
contrast than in pancakes, in the frame of ZA was emphasized in Arnold
{\it et al.}\cite{arn82}. This may explain why pancakes are not easily
seen in low {\it mass} resolution $N$--body simulations.
\section{Acknowledgements}

We thank Lev Kofman for a stimulating presentation which inspired this
study, and he as well as Phil Baringer, Ed Bertschinger, and Jerry
Ostriker provided useful discussions. Research support came from NSF
grant AST--9021414 (including Research Experiences for
Undergraduates), NASA grant NAGW--3832 and the National Center for
Supercomputing Applications. S.F. Shandarin acknowledges a University
of Kansas GRF 94 Grant.

\newpage

\section{FIGURE CAPTIONS}

FIG. 1 Three projections of the collapsed points (past the singular
stage) orthogonal to three principal axes of the initial deformation
tensor.

FIG. 2 Two thin slices (2 mesh units) approximately through the center
of the pancake orthogonal to (a) $z$--axis: $15\leq z\leq 17$; (b) $y$--axis
$10\leq y\leq 12$.

FIG. 3 The mass distribution in thin slices orthogonal to three
principal axis:

(a) $z$--axis: $15\leq z\leq 17$

(b) $y$--axis: $10\leq y\leq 12$, and

(c) $x$--axis: $112\leq x\leq 114$.
\newpage
\begin {table}
\centering
\caption{}
\label{}
\begin{tabular}{cccc}
Thickness & Width in & Length in \\
in mesh units & mesh units & mesh units \\
0.1 & 3.5 & 8 \\
0.5 & 5 & 16 \\
0.6 & 32 & 37 \\
0.8 & 7.5 & 48 \\
0.8 & 17 & 37 \\
\end{tabular}
\end{table}
\centerline{Information on the objects in the five simulations}

\end{document}